\documentclass[aps,amsmath,amssymb,prl,twocolumn,superscriptaddress]{revtex4}
\usepackage{bm}
\usepackage{epsfig}
\usepackage[usenames]{color}
\usepackage{graphicx}
\usepackage{ulem} 
\renewcommand{\paragraph}[1]{\textit{#1.---} }
\newcommand{\paper}{Letter }
\newcommand{\etal}{{\it et al.}}

\begin{document}

\title{Statistics of Deep Energy States in Coulomb Glasses}
\author{A. Glatz}\author{V.~M. Vinokur}
\affiliation{Materials Science Division, Argonne National
Laboratory, Argonne, IL 60439, USA}
\author{Y.~M. Galperin}
\affiliation{Department of Physics \& Centre of Advanced Materials
and Nanotechnology, University of Oslo, PO Box 1048 Blindern, 0316
Oslo,~Norway \\
and A. F. Ioffe
Physico-Technical Institute of Russian Academy of Sciences, 194021
St. Petersburg, Russia}
 \affiliation{Materials Science Division, Argonne
National Laboratory, Argonne, IL 60439, USA}

\date{\today}

\begin{abstract}
We study the statistics of local energy minima in the configuration
space and the energy relaxation due to activated hopping in a system
of interacting electrons in a random environment. The distribution
of the local minima is exponential, which is in agreement with
extreme value statistics considerations. The relaxation of the
system energy shows logarithmic time dependence reflecting 
the \emph{ultrametric} structure of the system.
\end{abstract}

\maketitle

Possible phases and the corresponding dynamic
behaviors of strongly correlated disordered systems remain one of
the central unresolved issues.  Doped semiconductors in the
insulating state are an exemplary system endowed with strong
long-range Coulomb interactions and strong disorder. It was
hypothesized that the combined action of both inflict the glassy
phase, resulting in the term \emph{electron- or Coulomb
glass}~\cite{davies}. Ever since, a Coulomb glass has been a subject of the intense
research, see~\cite{zvi}; however, despite the impressive
developments, a consistent picture of its nature is still not
available and the consensus on the interrelation between its
inherent glassy ingredients like Coulomb gap, slow dynamics and
aging, the nature of the low-lying metastable states and the
mechanisms of hopping transport is not yet achieved. A classic
qualitative derivation~\cite{ES}  of the Coulomb gap induced in the
single-particle density of states due to pair correlations remains
the top result in the field.

One of the main characteristics of the glassy state is the existence
of an infinite number of low-lying states (valleys in the rugged
free energy relief) separated by barriers growing infinitely in the
thermodynamic limit.  This picture emerged from the study of the
infinite range interaction model for spin glasses
(Sherrington-Kirkpatrick spin glass)~\cite{SK} and was later shown
to imply \textit{an ultrametric structure} of the configurational
space~\cite{MPSTV}.

There was important recent progress in relating the appearance of
the Coulomb gap in electron systems to the generic glassy behavior
~\cite{Dobros}, where a lattice model of spinless interacting
electrons in the presence of randomness was considered and a
nonlinear screening theory was formulated to tackle Coulomb gap
formation and its relation to glassy freezing.  The formation of the
glass phase with a large number of metastable states was identified
via the emergence of a replica symmetry breaking instability.
In~\cite{Ioffe}, the connection between the glass phase
characteristics and a Coulomb gap was argued, noticing that within
the locator-approximation the correlated electron system maps to the
Sherrington-Kirkpatrick spin glass.  Remarkably, recent experimental
studies of low-frequency resistance noise in silicon inversion
layers indicated glassy freezing and the presence of long-range
correlations, consistent with the hierarchical picture of glassy
dynamics~\cite{Jar}.  This gives strong support to  a hierarchical
structure of experimentally observed Coulomb glasses and calls for
further comprehensive study of hierarchical glassy systems.

In this \paper we demonstrate that the distribution of low-lying
energy states in systems with an ultrametric configurational space
obeys \textit{extreme value statistics} giving rise to an
exponential distribution for the extremely low energies,
$p\propto\exp(-E/E_0)$, where $E_0$ is a characteristic energy. The
exponential tail is verified numerically for the lattice model of a
disordered Coulomb system. We find that system's energy relaxes with
time following the $\log t$ law, as indeed expected for ultrametric
hierarchical structures~\cite{ogi}. This offers strong evidence that
electronic systems with long-range Coulomb interactions form a
Coulomb glass with exponential distribution of lowest energy states.
\begin{figure}[b]
\includegraphics[width=0.8\columnwidth,clip=true,viewport=.0cm 0.0cm
23.07cm 10.3cm]
{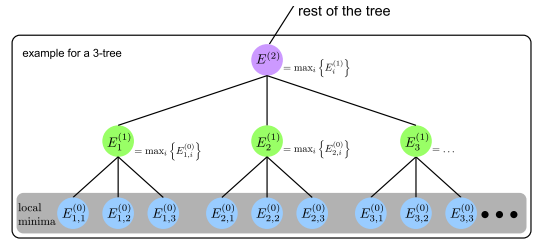}
\vspace*{-2mm}
\caption{Illustration of a 3-tree, an element of the hierarchical
tree representing the configurational space with ultrametric
topology.}\label{fig.tree}
\end{figure}

\paragraph{Exponential distribution} We start with the derivation of the
exponential statistics for the low-lying energy states of systems with
ultrametric topology of the configurational space.
This space can be represented graphically as a hierarchical tree
which is the union (or sum) of hierarchical trees with branching
numbers $n\in\mathbb{N}\backslash\{1\}$ ($n$-trees, see
Fig.~\ref{fig.tree}). The crown of the tree (leaves) corresponds to
a set of nearly equal energies that form an almost degenerate ground
state of the system. We consider now the subset of the
configurational space characterized by an $n$-tree and find the
distribution function of the corresponding lowest energy states.  It
is convenient to introduce positive variables $\{E\}$ measuring the
\textit{depths} of the potential wells. Starting with the lowest
hierarchical level (crown) consider one of the sibling states sets,
$\{E_i^{(0)};i\in\mathbb{N}^{\leq n}\}$, where the superscript $(0)$
marks the lowest hierarchy level. We define
$E_1^{(1)}\equiv\max_{1\leq i\leq n}\{E_i^{(0)}\}$ and assign it to
the branching point level $(1)$ generating the considered $0$-th set
of siblings. Now we have a collection of level-$(1)$ sibling
$n$-sets. Repeating the procedure, we obtain the next generation of
low-lying energies $\{E_i^{(2)}\}$ that constitute the assortment of
the second level sibling $n$-sets. The described iteration should
converge to a limiting distribution for the probability ${\cal
P}_n\left(E^{(j)}\right)$ for the lowest states at the $j$-level of
the hierarchy, as $j\to\infty$. This distribution must be stable
under the $\max$ operation. In other words if $E\equiv\max_i\{E_i\}$
then the probability distribution of the extrema $E$ is the same as
that of each member $E_i$ of the set. The distribution we seek
belongs to the class of so-called
extreme statistics~\cite{gumbel}.

To briefly summarize the theory of extreme distributions, we
consider a set of identically distributed independent random
variables $X_i$ , with $1\leq i\leq  n$ (in the present problem
these are all the low-lying energies within one set of sibling
states). Let $M_n = \max_{1\leq i\leq n}\{X_i\}$ and let ${\cal
P}_1(x)$ be the probability that any of the random variables $X_i$
is less than $x$. The probability that the largest value, $M_n$, of
the set $\{X_i;i\in\mathbb{N}^{\leq n}\}$ is less than $x$ is simply
the probability that all of the $X_i$'s fall short of $x$. Since the
variables are independent, this is given by
\begin{equation}
\label{prob_max} {\cal P}\left(\max_{1\leq i\leq n}\{X_i\}\leq
x\right)\equiv {\cal P}_n(x)=\left[{\cal P}_1(x)\right]^n.
\end{equation}
The central concept of the theory of extreme statistics is the
\textit{stability postulate}~\cite{gumbel}, stating that
after many iterations the distribution describing $M_n$ for a
sufficiently large set $\{X_i\}$ is the same as the distribution for
each variable $X_i$.  This requires that the probability ${\cal
P}_n(x)$ satisfied the functional equation $ {\cal P}_n(x)=\left[{\cal
  P}_n(a_nx+b_n)\right]^n$,
where $a_n$ and $b_n$ are positive constants.  It can be shown that,
under rather nonrestrictive assumptions about the initial
distribution of random variables $\{X_i\}$ there are only three
classes of distribution functions depending on its support.  In
particular, if $x\in\mathbb{R}$,
    \begin{equation}
         {\cal P}_n(x)=\exp[-n\exp(-x)]\, , \quad a_n=1, \  b_n=\ln(n),
    \end{equation}
and, therefore, the limiting form for a distribution of the lowest
energy states corresponding to an $n$-tree subset of the
configurational space is $ {\cal P}_n(E)=\exp[-n\exp(-E)]$,
where $E$ is measured in dimensionless energy units.

As mentioned before, the configurational space of our system endows
with an ultrametric structure which can be represented as an union of all possible $n$-trees.  Therefore the global
\textit{distribution density}
can be found as
\begin{equation} \label{glob-distr}
  p(E)=\int dn\,  w(n)\frac{d{\cal P}_n(E)}{dE},
\end{equation}
where $w(n)$ is the weight of the $n$-trees in the configurational
space. The weight $w(n)$ is determined by the physical properties of the
system in question.  In our case one observes that the elemental
process of generating two nearly degenerate energy states is the
pair electron exchange within a specially arranged compact 4-sites
aggregate~\cite{kozub+cm04}. Analogously, compact aggregates allowing the exchange of $n$ pairs,
generate the same number of sibling states of the corresponding
hierarchical trees. Thus $w(n)$ is nothing but the percolation
cluster size-distribution function: $w_n\propto n^{-s}$, where
$s\geq 2$. Plugging this weight into Eq.~(\ref{glob-distr}) one
arrives at the exponential density distribution for the low-lying
energies, $ p(E)\propto\exp(-E/E_0)$,
where $s$ is absorbed into the normalization energy $E_0$.

The extreme value distributions for extremely low-energy states
belongs to the so-called  were derived~\cite{Bouchaud} for several
systems and problems possessing one step replica symmetry breaking
solutions, such as the random energy model and decaying Burgers
turbulence. It was found that extremely low energy states of elastic
manifolds in random environment (modeling a rich variety of physical
systems including vortices in superconductors, charge density waves,
Wigner crystals, domain walls, and many others) also obey
exponential statistics resulting in  creep motion which is specific
for glassy states~\cite{VM}. Our results taken together with those
of~\cite{Bouchaud,VM} allow to arrive at the conclusion that the
exponential distribution of the extremely low-lying energy states,
the resulting creep response, and the related $1/f$
noise~\cite{burin+prb06} are generic properties of all strongly
correlated disordered systems possessing a fractal structure of
configurational space.
\begin{figure}[b]
\includegraphics[width=0.66\columnwidth]
{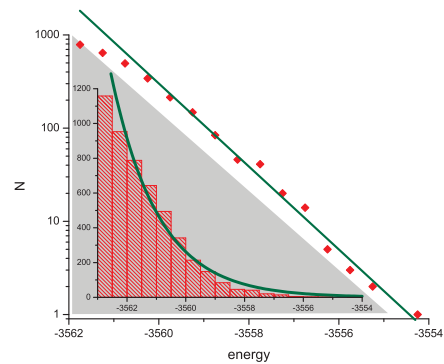}
\vspace*{-3mm}
\caption{Energy distribution of about $12000$ local
minimum states in half-log representation with linear fit. Linear
scale with exponential fit (triangular inset).}\label{fig.lmin}
\end{figure}

\paragraph{Model} Now we turn to numerical studies of the energy configurational
space. The lattice Coulomb model was studied numerically in the
past~\cite{baranovskii+jpc79,efros+jpc79,kogan-prb98}, the typical
system size was about $10^2$.  Optimized numerical procedures
enabled us to consider 100-times larger systems and analyze about
$10^5$ different energy minima. The Hamiltonian of a 2D Coulomb
lattice with disorder is
\begin{equation} \label{eq.model}
{\cal H}=\sum_i\varepsilon_i, \ \ \varepsilon_i=U \alpha_i n_i+\frac{e^2}{2}\sum_{j\neq
i}\frac{(n_i-\nu)(n_j-\nu)}{r_{ij}}\,,
\end{equation}
where $i$ and $j$ label the lattice sites, $r_{ij}=|{\bf r}_i-{\bf
r}_j|$, $U\alpha_i$ with $\alpha_i\in [-1;1]$ define the quenched
uniformly distributed random site energies, and $n_i$ are the
occupation numbers (values $0$ or $1$) with average $\nu$.

The simulations were performed on the lattices with typically
$N=100^2$ sites with periodic boundary conditions and the filling
factor $\nu=0.5$. We took the disorder spread, $U$, be equal to the
Coulomb interaction at the distance of the lattice constant.
Consequently, $U$, the lattice constant, and $e$ were set to unity;
vectors $\textbf{r}_i$ label square lattice sites. The behavior of
the system is governed by the interplay between disorder and Coulomb
interactions.  At very weak disorder the system is crystal-like;
increase in $U$ brings glassy behavior.

\paragraph{Local minima}
To track down local minima of the system free energy, we choose an
arbitrary initial distribution of electrons and perform sequential
transfer of electrons between the adjacent sites (of course such an
elemental hop can be realized only if the target neighboring site is
vacant). At each step the total Coulomb energy of the system is
calculated using Eq.(\ref{eq.model}).  If any nearest site hop gives
rise to the increase in total energy, as compared to its value at
previous step, the state and the corresponding energy is marked as
energy minimum. In subsequent iterations one finds transfers that
decrease total energy executing thus dissfusion of the system into
more and more deep energy valleys. Note that this algorithm lists
\textit{all} the possible low-lying states.  To ensure evolving the
system into true low lying minima we were starting collection
statistics after reaching the states where at least $10^5$ random
hops were needed to go from one local minimum to another.
Shown in Fig.~\ref{fig.lmin} is the obtained distribution of
energies corresponding to local minima; one sees that the
distribution tail obeys the exponential law, $N_{\text{min}}(E)\sim
e^{-E/E_0}$.
\begin{figure}[t]
\centering
\includegraphics[width=0.8\columnwidth]
{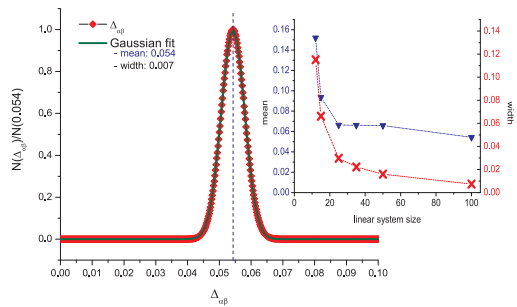}
\vspace*{-3mm}
\caption{Distribution of the difference of the local minima
$\Delta_{\alpha\beta}$ ($0$ for identical states, $1$ for
complementary ones). The distribution is fitted by a Gaussian curve,
with mean at $0.054$ and width $0.007$. $10^7$ overlap integrals are
calculated. The inset shows the dependence of the mean and
width of the distribution on system size.}\label{fig.overlap}
\end{figure}

For all the computed minima we define the normalized site occupation
number difference $\Delta_{\alpha\beta}=
N^{-1}\sum_i|n^{\alpha}_i-n^{\beta}_i|$ corresponding to all pairs
of minima $\{\alpha,\beta\}$. The distribution of
$\Delta_{\alpha\beta}$, i.~e., the fraction of the system which has
different occupation numbers for two different minima is shown in
Fig.~\ref{fig.overlap}. To drive our system (the filling factor
 $0.5$) from one minimum to another, about $5.4$\% of the electrons
has to be transferred to different lattice sites. This shows that
the (pseudo-)ground state of our system is indeed highly degenerate:
it is enough to move around only a few electrons in order to get
into another energetically close local minimum. This well
illustrates the ultrametric structure of the configurational space.

The $\Delta_{\alpha\beta}$ distribution is also calculated for
smaller systems with the same filling factor. For system sizes above
$\approx 25^2$ the fraction of the electrons different between
minima and the width of the distribution stay essentially the same.
Both increase upon lowering the system size; see inset in
Fig.~\ref{fig.overlap}. For very small systems (below $15^2$) the
distribution is not Gaussian anymore since there is only a small
number of possible local minima. This indicates that small systems
do not have an ultrametric structure and therefore do not represent
physical systems.

\begin{figure}[t]
\centering
\includegraphics[width=0.75\columnwidth]
{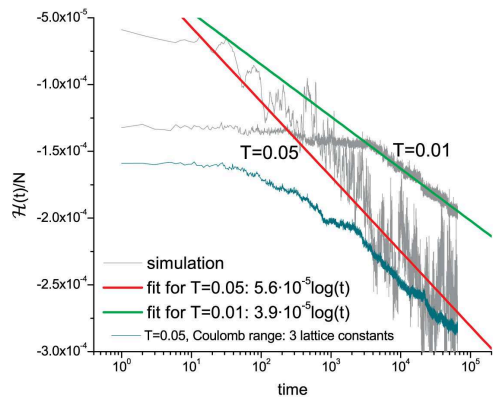}
\vspace*{-3mm}
\caption{Energy relaxation of the system energy for two different
temperatures. In the two upper plots logarithmic fits are shown. At
lower temperatures ($T=0.01$) the initial ``flat'' region is due to
a very long equilibration phase of the system to reach the steady
state from the local minimum. Even after averaging over several time
series the fluctuations at higher temperatures are still large. At
even higher temperatures the fluctuations dominate and the
logarithmic behavior cannot be recognized since due to computational
limits no more averaging can be done in reasonable time. The lower
graph shows the energy relaxation for the case when the Coulomb
interaction range is artificially limited to 3 lattice sites at
$T=0.05$. One sees (1) that the behavior deviates from the log
behavior and (2) the amplitude of the fluctuations are much smaller
as for the case with full Coulomb interaction.}\label{fig.relax}
\end{figure}

\paragraph{Energy relaxation} Next we analyze the relaxation of the system
at finite temperatures. Due to the ultrametric structure of the
system the relaxation can be described as diffusion or random walk
on the tree representing the structure~\cite{ogi}
(Fig.~\ref{fig.tree}). Starting from a local minimum, nearest
neighbor hopping with Boltzmann factors
$e^{-\Delta\varepsilon_{ij}/T}$ is applied to all particles in the
system - random permutations are used to address the lattice
positions in one time step. The typical energy difference
$\Delta\varepsilon_{ij}$ of two adjacent lattice sites $i$ and $j$
is of order one in the used dimensionless units.

In Fig.~\ref{fig.relax} the time evolution of the system energy is
shown for two different low temperatures. Since the relaxation is
very slow and the fluctuations are significant (even after averaging
over several time series) and furthermore increase with temperature,
a conclusive discrimination of the behavior is not possible.
However, it can be well fitted by a logarithmic law in agreement
with the result of~\cite{ogi}.  This lends an additional support to
the hypothesis of the glassy nature of long-range interacting,
disordered Coulomb systems. Previously the time evolution in the
Coulomb lattice model was studied via Monte-Carlo methods, see
e.g.~\cite{efros+ssc95, shklovskii-prb03,tsigankov+prb03}.
In~\cite{tsigankov+prb03} the energy relaxation observed via
evolution of the system after adding a few electrons showed $1/t$-behavior.

Finally, we investigated the relation between the relaxation of the
system and electron correlations reducing artificially the screening
length of the Coulomb interaction (in all other simulations the
range is of order of the system size). In this case the relaxation
deviates from the $\log t$ law (see lower curve in
Fig.~\ref{fig.relax}) indicating the crucial role of long-range
interactions in forming Coulomb glass.

\paragraph{Discussion and conclusions}
In conclusion, we have derived that in a system endowed with an
ultrametric configurational space the local energy minima follow
extreme value statistics resulting in an exponential distribution of
the low-lying energy states.  Since in the most strongly correlated
disordered systems the energy landscape is expected to be
characterized by only one energy parameter, one can assume that
barriers separating the low-lying states also obey the same
statistics. This suggests that the nonlinear creep dynamic response
and the related $1/f$ noise are generic properties of all strongly
correlated disordered systems.

Our numerical results for a disordered system with long-range
Coulomb interaction -- including the exponential distribution of the
low-lying energy states and the logarithmic time relaxation of the
system towards its ground state -- offer strong evidence towards the
existence of a Coulomb or electronic glass and relates the formation
of the glassiness with long-range correlations in Coulomb systems.
This further explains why pronounced glassy properties like aging
and memory effects as well as apparently unbounded $1/f$ noise are
observed in electronic systems with profound long-range
electron-electron interactions. Of course the specific manifestation
of the energy statistics in the observable quantities requires
concrete consideration of the peculiarities of the physical system
involved and goes beyond the scope of this work. In particular, one
of the most intriguing questions concerning the behavior and the
nature of the noise in various disordered electronic systems will be
the subject of a forthcoming publication.

\paragraph{Acknowledgements} We thank A. L. Burin and V.~I.~Kozub for
useful and encouraging discussions. This work was supported by the
U.S. Department of Energy Office of Science under the Contract No.
DE-AC02-06CH11357. AG acknowledges support by the Deutsche
Forschungsgemeinschaft.

\end{document}